\documentclass[fleqn,usenatbib]{mnras}

\usepackage{newtxtext,newtxmath}

\usepackage[T1]{fontenc}

\DeclareRobustCommand{\VAN}[3]{#2}
\let\VANthebibliography\thebibliography
\def\thebibliography{\DeclareRobustCommand{\VAN}[3]{##3}\VANthebibliography}

\usepackage{graphicx}	
\usepackage{amsmath}	

\title[The formation of cores in galaxies across cosmic time]{The formation of cores in galaxies across cosmic time - the existence of cores is not in tension with the $\Lambda$CDM paradigm}

\author[R. A. Jackson et al.]
{R. A. Jackson,$^{1,2,3}$\thanks{E-mail: ryanjackson@uvic.ca}
S. Kaviraj,$^{2}$ S. K. Yi,$^{3}$ S. Peirani,$^{4,5}$ Y. Dubois,$^{5}$ G. Martin,$^{6,7,8}$\newauthor 
J. E. G. Devriendt,$^{9}$ A. Slyz,$^{9}$
C. Pichon,$^{5,10}$
M. Volonteri,$^{5}$
T. Kimm$^{3}$ and
K. Kraljic$^{11}$
\\
$^{1}$Department of Physics and Astronomy, University of Victoria, BC V8P 5C2, Canada\\
$^{2}$Centre for Astrophysics Research, School of Physics, Astronomy and Mathematics, University of Hertfordshire, Hatfield, AL10 9AB, UK\\
$^{3}$Department of Astronomy and Yonsei University Observatory, Yonsei University, Seoul 03722, Republic of Korea\\
$^{4}$Observatoire de la C$\hat{\rm{o}}$te d'Azur, CNRS, Laboratoire Lagrange, Bd de l'Observatoire, Universit\'e C$\hat{\rm{o}}$te d'Azur, CS 34229, 06304 Nice Cedex 4, France \\
$^{5}$Institut d'Astrophysique de Paris, Sorbonne Universit\'es, UMPC Univ Paris 06 et CNRS, UMP 7095, 98 bis bd Arago, 75014 Paris, France\\
$^{6}$School of Physics and Astronomy, University of Nottingham, University Park, Nottingham NG7 2RD, UK\\
$^{7}$Korea Astronomy and Space Science Institute, 776 Daedeokdae-ro, Yuseong-gu, Daejeon 34055, Korea\\
$^{8}$Steward Observatory, University of Arizona, 933 N. Cherry Ave, Tucson, AZ 85719, USA\\
$^{9}$Dept of Physics, University of Oxford, Keble Road, Oxford OX1 3RH UK\\
$^{10}$School of Physics, Korea Institute for Advanced Study (KIAS), 85 Hoegiro, Dongdaemun-gu, Seoul, 02455, Republic of Korea\\
$^{11}$Observatoire Astronomique de Strasbourg, Universit\'e de Strasbourg, CNRS, UMR 7550, F-67000 Strasbourg, France\\
}

\pubyear{2015}

\begin{document}
\label{firstpage}
\pagerange{\pageref{firstpage}--\pageref{lastpage}}
\maketitle

\begin{abstract}
The `core-cusp' problem is considered a key challenge to the $\Lambda$CDM paradigm. Halos in dark matter only simulations exhibit `cuspy' profiles, where density continuously increases towards the centre. However, the dark matter profiles of many observed galaxies (particularly in the dwarf regime) deviate strongly from this prediction, with much flatter central regions (`cores'). We use \texttt{NewHorizon} (NH), a hydrodynamical cosmological simulation, to investigate core formation, using a statistically significant number of galaxies in a cosmological volume. Halos containing galaxies in the upper (M$_{\star}$ $\ge$ 10$^{10.2}$ M$_{\odot}$) and lower (M$_{\star}$ $\le$ 10$^{8}$ M$_{\odot}$) ends of the stellar mass distribution contain cusps. However, halos containing galaxies with intermediate (10$^{8}$ M$_{\odot}$ $\le$ M$_{\star}$ $\le$ 10$^{10.2}$ M$_{\odot}$) stellar masses are generally cored, with typical halo masses between 10$^{10.2}$ M$_{\odot}$ and 10$^{11.5}$ M$_{\odot}$. Cores form through supernova-driven gas removal from halo centres, which alters the central gravitational potential, inducing dark matter to migrate to larger radii. While all massive (M$_{\star}$ $\ge$ 10$^{9.5}$ M$_{\odot}$) galaxies undergo a cored-phase, in some cases cores can be removed and cusps reformed. This happens if a galaxy undergoes sustained star formation at high redshift, which results in stars (which, unlike the gas, cannot be removed by baryonic feedback) dominating the central gravitational potential. After cosmic star formation peaks, the number of cores, and the mass of the halos they are formed in, remain constant, indicating that cores are being routinely formed over cosmic time after a threshold halo mass is reached. The existence of cores is, therefore, not in tension with the standard paradigm.


\end{abstract}




\section{Introduction}

The study of dwarf galaxies is a key area of research in the current astronomical landscape. Dwarfs are important because they dominate the galaxy number density, across all environments and at all epochs \citep{Wright2017,Martin2019,Jackson2021b}. Given their low stellar masses and high dark matter (DM) content \citep{DiCintio2017,Chan2018,Jackson2021a}, dwarfs also provide a unique laboratory for testing our current understanding of DM and the $\Lambda$CDM structure formation paradigm \citep{Davis2022}. However, dwarf galaxies remain difficult to study, both observationally and theoretically, due to the shallow detection limits of recent wide-area surveys such as the SDSS \citep{Alam2015} and the relatively low resolution of large volume simulations (e.g. Horizon-AGN \citep{Dubois2014,Kaviraj2017}, EAGLE \citep{Schaye2015}, Illustris \citep{Vogelsberger2014}, APOSTLE \citep{Sawala2016}, Simba \citep{Dave2019}, IllustrisTNG \citep{Nelson2019}). It is, therefore, unsurprising that many tensions between observations and $\Lambda$CDM predictions are found in the dwarf regime \citep[e.g.][]{Bullock2017,Sales2022}. 

One key tension is the `core-cusp' problem \citep{Flores1994,Moore1994}, which describes the observed discrepancy between the measured dark matter (DM) density slopes in cosmological simulations and those inferred from the rotation curves of local dwarf galaxies \citep{Blok2010}. Simulations of cold DM predict a cuspy DM density profile, where density increases towards the centre of the halo \citep[$\rho\propto$ r$^{-1}$, e.g.][]{White1978,Springel2008}. This is commonly described by a Navarro-Frenk-White (NFW) profile \citep{Navarro1996b,Navarro1997}, although other fits to this profile have also been suggested \citep[e.g.][]{Einasto1965}. In contrast, observations of dwarf galaxies often find slowly rising rotation curves which would indicate that the profiles instead have a much shallower (or even flat) density slope (i.e. a core) in their inner regions with $\rho\propto$ r$^{\sim0}$ \citep[e.g.][]{Burkert1995,deBlok2001,Gilmore2007,deBlok2008,KuziodeNaray2008,Kormendy2009,Oh2011,Oh2015,Lelli2016}.

This discrepancy has three main suggested solutions. First, it is possible that the discrepancy is driven by systematic errors (or unknown uncertainties) in the measurement of dwarf galaxy rotation curves. Given the difficulty in observing dwarf galaxies and extracting accurate rotation curves, this is a plausible solution. Many sources of error (such as the failure to account for non-circular motions) have been discussed and studied using simulations \citep{Strigari2017,Genina2018,Harvey2018,Oman2019,Santos-Santos2020,Roper2023,Downing2023}. Another solution involves modifications to the $\Lambda$CDM model, such as invoking warm DM \citep{Dodelson1994,Bode2001} or self-interacting DM \citep[SIDM, ][]{Yoshida2000,Spergel2000,Vogelsberger2012,Rocha2013,Tulin2018}. SIDM has, in particular, been recently proposed as a viable alternative which solves the core-cusp problem \citep{Burger2019,Burger2022}. Finally, there is a potential solution within $\Lambda$CDM itself, which relies on the effect that baryons and baryonic processes may themselves have on DM density profiles. 

Baryonic feedback processes that operate within galaxies, driven by active galactic nuclei (AGN) and/or supernovae (SN), are capable of redistributing mass within the DM halos, which can then alter the shape of the DM density profile via noise driven orbital shuffling \citep{Weinberg2001}. Using the Horizon-AGN simulation, \citet{Peirani2017} have shown that AGN can affect the shape of a halo's DM density profile in massive galaxies. They show that DM density profiles in a Horizon-AGN run that includes baryons are initially steeper than their DM-only counterpart. However, by $z\sim1.5$ AGN activity is able to flatten these profiles. After AGN activity ceases the profile then returns to being a cusp. Other work has also shown that AGN can create cores in massive galaxies at high redshift \citep{Dekel2021,Li2023}. More commonly in dwarf galaxies (where many cores are found in observations), supernova (SN) feedback is known to similarly be able to remove gas from the central regions of galaxies \citep{Dubois2008,DiCintio2017,Chan2018,Jackson2021b}. 

Many theoretical studies have shown that SN feedback can create cores via violent or repeated bursts of star formation \citep{Dekel1986,Navarro1996,Gnedin2002,Read2005,Mashchenko2008,Pontzen2012,Garrison-Kimmel2013,Teyssier2013,DiCintio2014,Chan2015,Dutton2016,Tollet2016,Fitts2017,Freundlich2020,Lazar2020,Burger2021}. One method by which this can occur, and which is invoked to explain cores in several simulations, is via the adiabatic change of the DM potential. Briefly, the gravitational potential of the DM halo is perturbed by SN-induced fluctuations which stirs the gas. DM particles on average gain energy and migrate to larger orbits due to these stochastic perturbations which resonate with their orbital frequencies, leading to lower central densities and a cored DM density profile. 

Since the early attempts to model galaxies and DM halos with N-body simulations \citep[e.g.][]{Somerville1999,Cole2000,Benson2003,Bower2006,Croton2006}, computational power has increased exponentially. We are now able to run large, hydrodynamical simulations in cosmological volumes, which can model baryons and DM self-consistently, allowing us to test the interplay between them and quantify the influence of baryonic processes on the DM. Different cosmological simulations, that use a variety of underlying codes and subgrid models, have subsequently attempted to address the core-cusp problem. Those which adopt a zoom-in technique (resulting in generally higher resolution but lacking a statistical sample) have had success in both recreating the broad properties of galaxies across a range of stellar masses and creating DM density profiles with both cores and cusps. These simulations \citep[e.g.][]{DiCintio2014,Tollet2016,Chan2015,Fitts2017,Lazar2020} consistently find that SN feedback is capable of driving gas outflows that alter the gravitational potential in such a way as to form DM cores in dwarf galaxies. They have also shown that cores form more efficiently in a relatively narrow range of stellar or halo mass and that the inner slope of dark matter haloes may be halo mass dependent. 

Conversely, large-volume simulations such as Illustris \citep{Vogelsberger2014}, EAGLE \citep{Schaye2015} and APOSTLE \citep{Sawala2016}, which reproduce the properties and scaling relations that govern massive galaxies well, have been unable to reproduce cored DM density profiles in their fiducial runs, regardless of the stellar or halo mass of the galaxies in question \citep{Schaller2015,Benitez-Llambay2019,Chua2019}. However, re-simulations using the EAGLE model have managed to produce a diversity of DM density profiles when adjusting the gas density threshold required for star formation, albeit for smaller samples of galaxies \citep{Benitez-Llambay2019}. 

This inability to form cores in large volume simulations is attributed to two causes: the star formation (SF) density threshold of the gas and the modelling of the inter-stellar medium (ISM). These parameters can often be linked, with a lack of spatial resolution in simulations leading to approximations in the modelling of the ISM, and a low SF density threshold leading to a lack of cores \citep{Oman2015,Schaller2015,Bose2019}. Indeed, many subsequent studies have shown that efficient core formation relies on the presence of dense gas (and therefore a high SF threshold), as well as bursty SF, in order to perturb the gravitational potential sufficiently to move DM particles to larger radii \citep{Governato2010,Teyssier2013,DiCintio2014,Benitez-Llambay2019,Jahn2021}. Therefore, a simulation which is both capable of resolving the small-scale processes within galaxies and offers a cosmological volume containing a statistically large sample of dwarf galaxies is needed to truly address the core-cusp problem. 

\begin{figure*}
\centering
\includegraphics[width=0.9\textwidth]{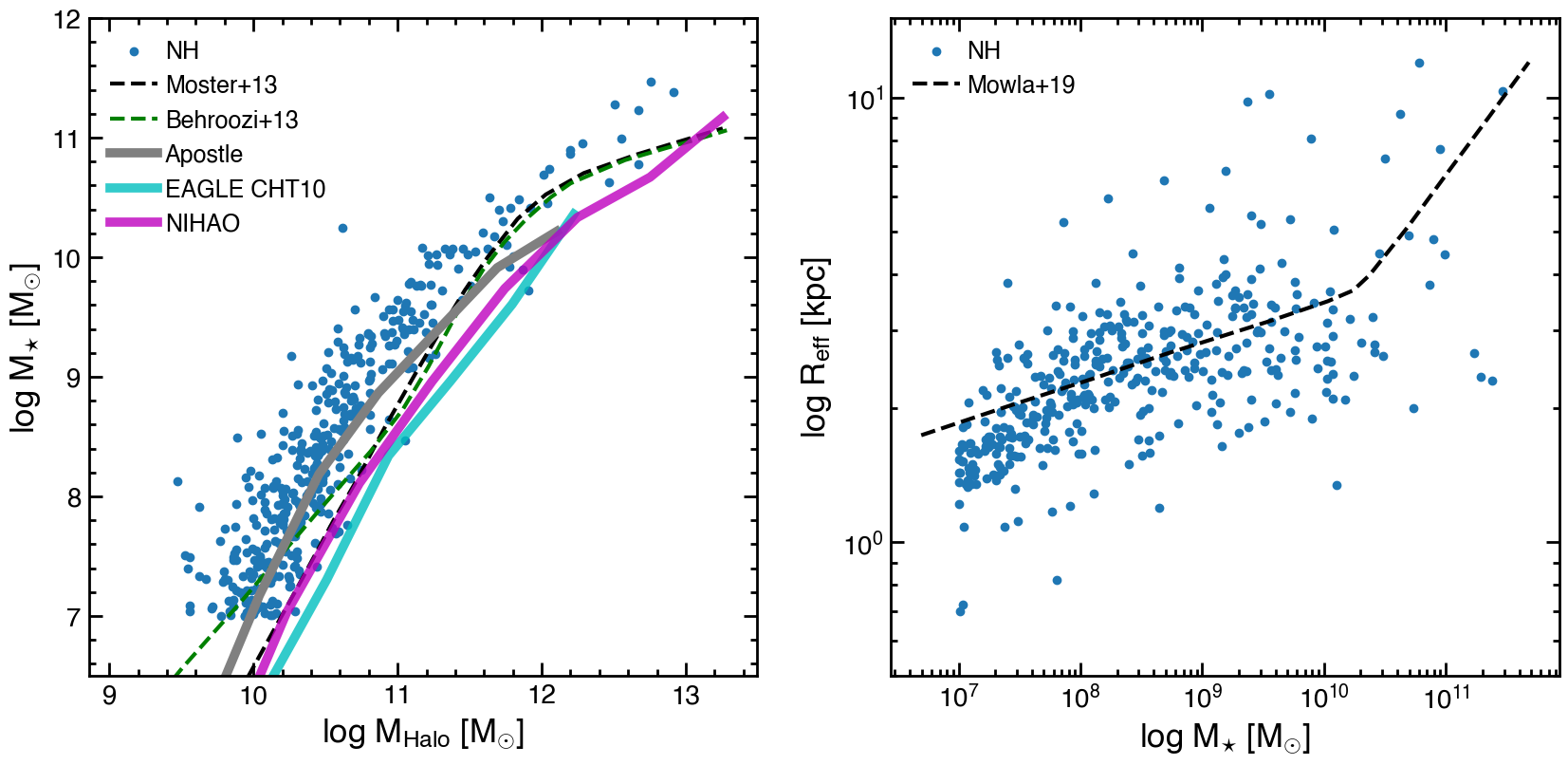} \caption{Left: M$_{\rm{Halo}}$ vs M$_{\star}$ for the \texttt{NewHorizon} galaxy sample used in this study (blue circles). We compare to the abundance matching studies from \citet{Behroozi2013} and \citet{Moster2013} and other simulations: APOSTLE \citep{Sawala2016}, EAGLE CHT10 \citep{Benitez-Llambay2019} and NIHAO \citep{Wang2015}. On average, NH halos contain more stellar mass than other simulations, with the observational data lying between NH and the other simulations. Right: M$_{\star}$ vs R$_{\rm{eff}}$ for the NH galaxy sample (blue circles). We also show the best fit observational results from \citet{Mowla2019} ($z=0.37$). The sizes of NH galaxies are well-matched to the observations, particularly in the stellar mass range of interest in this study. For more details of how NH matches to observational results, see \citet{Dubois2021}.}
\label{fig:sim_comp}
\end{figure*}

In this study we use the \texttt{NewHorizon} hydrodynamical cosmological simulation \citep{Dubois2021}, to study the DM density profiles of galaxies across all stellar masses and in a range of environments (from the field to large groups). \texttt{NewHorizon} is a zoom-in of an average density region taken from the larger Horizon-AGN simulation \citep[][]{Dubois2014,Kaviraj2017}, which has a volume of (142 cMpc)$^{3}$. \texttt{NewHorizon} offers a maximum spatial resolution of 34 pc and mass resolutions of $\sim$10$^4$ M$_{\odot}$ and $\sim$10$^6$ M$_{\odot}$ in the stars and DM respectively. In addition, it has a timestep resolution of 15 Myr. This makes it an ideal tool to study the inner DM density profiles of galaxies, particularly in the dwarf regime where cores are believed to be present \citep[e.g.][]{Jackson2021a,Jackson2021b,Martin2021}, whilst also offering a statistically significant sample of galaxies due to its large volume. 

This paper is structured as follows. In Section \ref{sec:NH} we briefly describe the \texttt{NewHorizon} simulation and in Section \ref{sec:CC} we show the broad properties of the DM density profiles in NH galaxies.In Section \ref{sec:CCevo} we study how cores and cusps form in the simulation. We summarise our findings in Section \ref{sec:summary}.

\begin{figure*}
\centering
\includegraphics[width=0.9\textwidth]{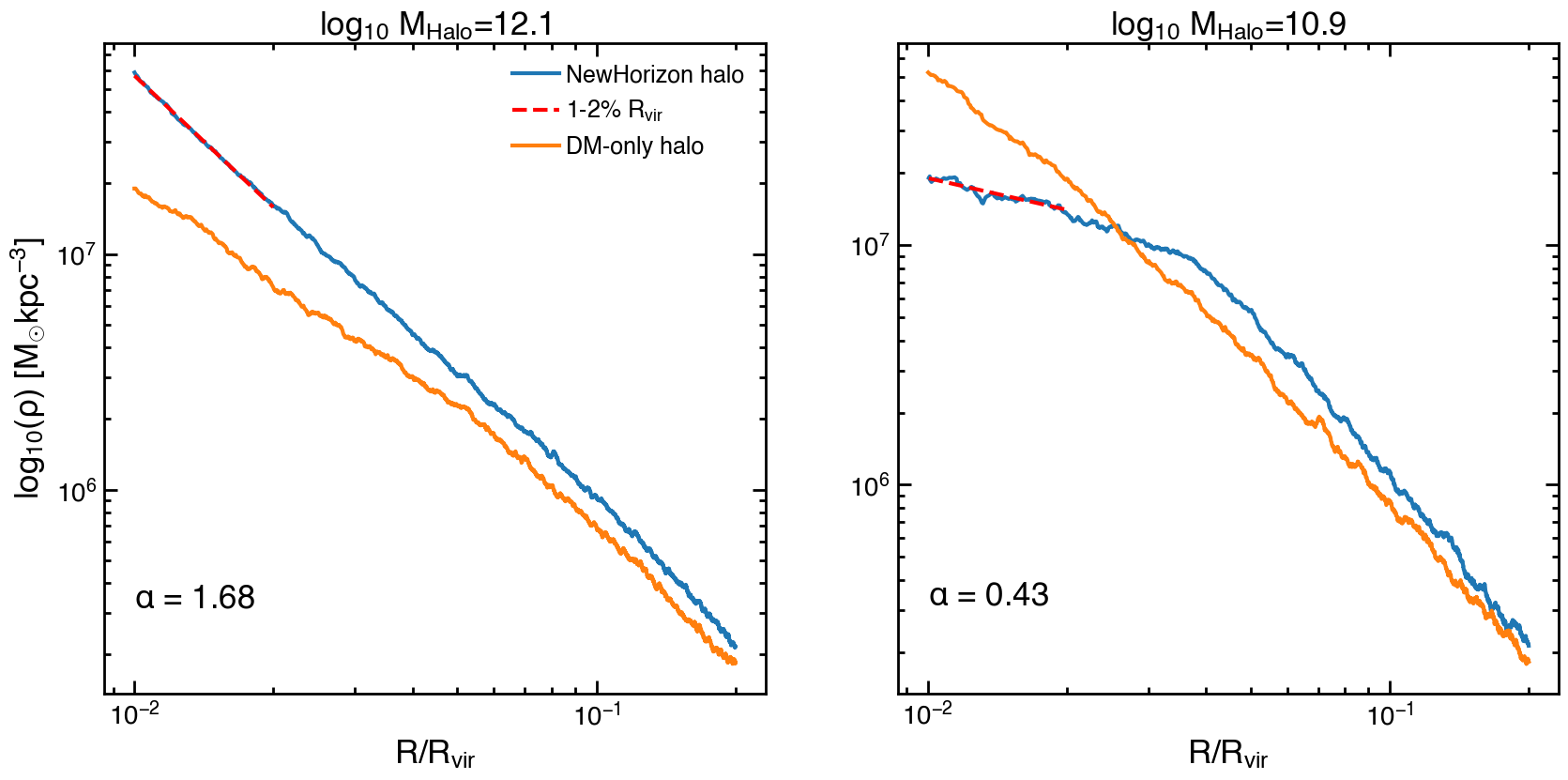}
\caption{The DM density profiles of two galaxies in the NewHorizon simulation (blue lines) and their counterparts in the NH DM-only run (orange lines) at $z=0.18$. The left panel shows an example of an NFW-like halo, where the density continues to increase towards the centre, with a slope of 1.68. In the DM-only run, the central regions have lower densities compared to the NewHorizon run, in which the presence of baryons in these central regions further deepens the potential. The right panel shows an example of a core-like halo with a slope of 0.43, in which the DM has been redistributed from the centre to larger radii in the NewHorizon run compared to the DM-only run. The NewHorizon simulation is therefore able to produce both cuspy and core-like profiles in the presence of baryons. In both examples, we indicate the region within which the slope is calculated using the dashed red line.}
\label{fig:corecusp_ex}
\end{figure*}

\section{Simulation}
\label{sec:NH}

The \texttt{NewHorizon} cosmological, hydrodynamical simulation \citep[][]{Dubois2021}, is a high-resolution simulation, produced using a zoom-in of a region of the Horizon-AGN simulation. Horizon-AGN employs the adaptive mesh refinement code RAMSES \citep{Teyssier2002} and utilises a grid that simulates a (142 cMpc)$^{3}$ volume (achieving a spatial resolution of 1 kpc), using 1024$^3$ uniformly-distributed cubic cells that have a constant mass resolution, via the \texttt{MPGrafic} \citep{Prunet2008} code. 

For \texttt{NewHorizon}, we resample this grid at higher resolution (using 4096$^3$ uniformly-distributed cubic cells), with the same cosmology as that used in Horizon-AGN ($\Omega_m=0.272$, $\Omega_b=0.0455$, $\Omega_{\Lambda}=0.728$, H$_0=70.4$ km s$^{-1}$ Mpc$^{-1}$ and $n_s=0.967$; \citet{Komatsu2011}). The high-resolution volume occupied by \texttt{NewHorizon} is a sphere which has a radius of 10 comoving Mpc, centred on a region of average density within Horizon-AGN. \texttt{NewHorizon} has a DM mass resolution of 10$^6$ M$_{\odot}$ (compared to 8$\times$10$^7$ M$_{\odot}$ in Horizon-AGN), stellar mass resolution of 10$^4$ M$_{\odot}$ (compared to 2$\times$10$^6$ M$_{\odot}$ in Horizon-AGN) and a maximum spatial resolution of 34 pc (compared to 1 kpc in Horizon-AGN)\footnote{The gravitational force softening is equal to the local grid size.}. The simulation has been performed down to $z\sim0.18$. 


\subsection{Star formation and stellar feedback} 

\texttt{NewHorizon} employs gas cooling via primordial Hydrogen and Helium, which is gradually enriched by metals produced by stellar evolution \citep{Sutherland1993,Rosen1995} allowing for metal cooling. An ambient UV background is switched on after the Universe is re-ionized at $z=10$ \citep{Haardt1996}. Star formation is assumed to take place in gas that has a hydrogen number density greater than $n_{\rm{H}}>$10 H cm$^{-3}$ and a temperature lower than 2$\times$10$^4$ K, following a Schmidt relation \citep{schmidt1959}. The star-formation efficiency is dependent on the local turbulent Mach number and the virial parameter $\alpha_{\rm vir}=2E_k$/|$E_g$|, where E$_k$ is the turbulent kinetic energy of the gas and E$_g$ is the gas gravitational binding energy \citep{Kimm2017}. The probability of forming a star particle of mass M$_{*,res}$=10$^4$ M$_{\odot}$ is drawn at each time step using a Poissonian sampling method, as described in \citet{Rasera2006}. 

Each star particle represents a coeval stellar population, with 31 per cent of the stellar mass of this population (corresponding to stars more massive than 6 M$_{\odot}$) assumed to explode as Type II
supernovae, 5 Myr after its birth. This fraction is calculated using a Chabrier initial mass function \citep{Chabrier2005}, with upper and lower mass limits of 150 M$_{\odot}$ and 0.1 M$_{\odot}$. Supernova feedback takes the form of both energy and momentum, with the final radial momentum capturing the snowplough phase of the expansion \citep{Kimm2014}. The initial energy of each supernova is 10$^{51}$ erg and the supernova has a progenitor mass of 10 M$_{\odot}$. Pre-heating of the ambient gas by ultraviolet radiation from young O and B stars is included by augmenting the final radial momentum from supernovae following \citet{Geen2015}.


\subsection{Supermassive black holes and black-hole feedback}

Supermassive black holes (SMBHs) are modelled as sink particles. These accrete gas and impart feedback to their ambient medium via a fraction of the rest-mass energy of the accreted material. SMBHs are allowed to form in regions that have gas densities larger than the threshold of star formation, with a seed mass of 10$^4$ M$_{\odot}$. New SMBHs do not form at distances less than 50 kpc from other existing black holes. A dynamical gas drag force is applied to the SMBHs \citep{Ostriker1999} and two SMBHs are allowed to merge if the distance between them is smaller than 4 times the cell size, and if the kinetic energy of the binary is less than its binding energy.

The Bondi-Hoyle-Lyttleton accretion rate determines the accretion rate on to SMBHs, with a value capped at Eddington \citep{hoyle1939,Bondi1944}. The SMBHs release energy back into the ambient gas via a thermal quasar mode and a jet radio mode, when accretion rates are above and below 1 per cent of the Eddington rate respectively \citep{Dubois2012}. The spins of these SMBHs are evolved self-consistently through gas accretion in the quasar mode and coalescence of black hole binaries \citep{Dubois2014a}, which modifies the radiative efficiencies of the accretion flow (following the models of thin Shakura \& Sunyaev accretion discs) and the corresponding Eddington accretion rate, mass-energy conversion, and bolometric luminosity of the quasar mode \citep{Shakura1973}. The quasar mode imparts 15 per cent of the bolometric luminosity as thermal energy into the surrounding gas. The radio mode employs a spin-dependent variable efficiency and spin up and spin down rates that follow the simulations of magnetically-choked accretion discs \citep[see e.g.][]{Mckinney2012}. 


\subsection{Identification of galaxies and merger trees}

We use the AdaptaHOP algorithm to identify DM halos \citep{Aubert2004,Tweed2009}. AdaptaHOP efficiently removes subhalos from principal structures and keeps track of the fractional number of low-resolution DM particles within the virial radius of the halo in question. We identify galaxies in a similar fashion, using the HOP structure finder applied directly to star particles \citep{Eisenstein1998}. The difference between HOP and AdaptaHOP lies in the fact that HOP does not remove substructures from the main structure, since this can result in star-forming clumps being removed from galaxies. We then produce merger trees for each galaxy in the final snapshot at $z\sim0.18$, with an average timestep of $\sim$ 15 Myr, enabling us to track the main progenitor of every galaxy with high temporal resolution.

Given that \texttt{NewHorizon} is a high resolution zoom of Horizon-AGN, it is worth considering the DM purity of galaxies, since higher-mass DM particles may enter the high resolution region of \texttt{NewHorizon} from the surrounding lower-resolution regions. Given the large mass difference, these DM particles may interact with galaxies they are passing through in unusual ways. The vast majority of galaxies affected by low DM purity exist at the outer edge of the \texttt{NewHorizon} sphere. The galaxies studied in this paper all have DM halos with a purity of 100 per cent at the final snapshot.

\section{Cores and cusps in the galaxy population}
\label{sec:CC}

\begin{figure}
\centering
\includegraphics[width=\columnwidth]{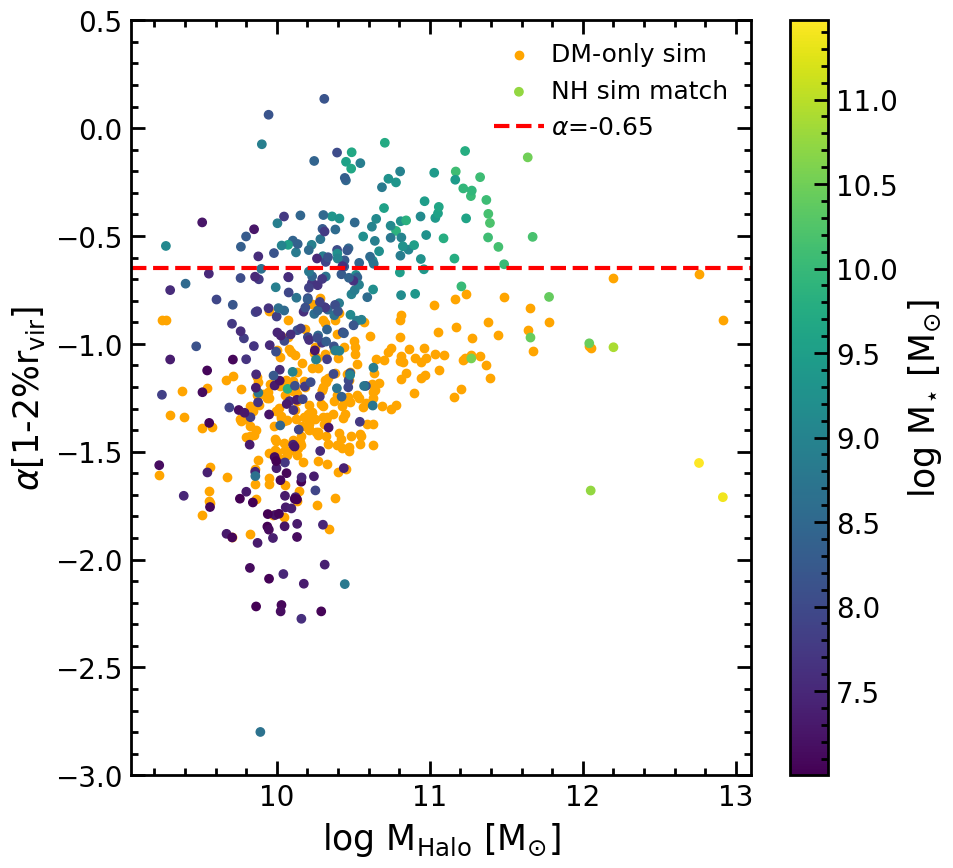}
\caption{The slope of the density profile for galaxies in the full NH (i.e. DM + baryons) simulation, and its DM-only counterpart, as a function of halo mass. Galaxies from the full run are colour-coded by their stellar mass. We show our definition of a cored halo ($\alpha=-0.65$) using a dashed red line. Galaxies are generally cusps at the lowest halo masses, whilst at intermediate masses cores form. As halo masses continue to grow, profiles return to cusps at the highest masses. There is a transition halo mass (M$_{\rm{halo}}$ $\sim$ 10$^{10.2}$M$_{\odot}$) where the profile slope generally flattens, compared to the DM-only run suggesting that baryonic processes are key for the creation of cores.}
\label{fig:NH_DM-only}
\end{figure}

\begin{figure*}
\centering
\includegraphics[width=\textwidth]{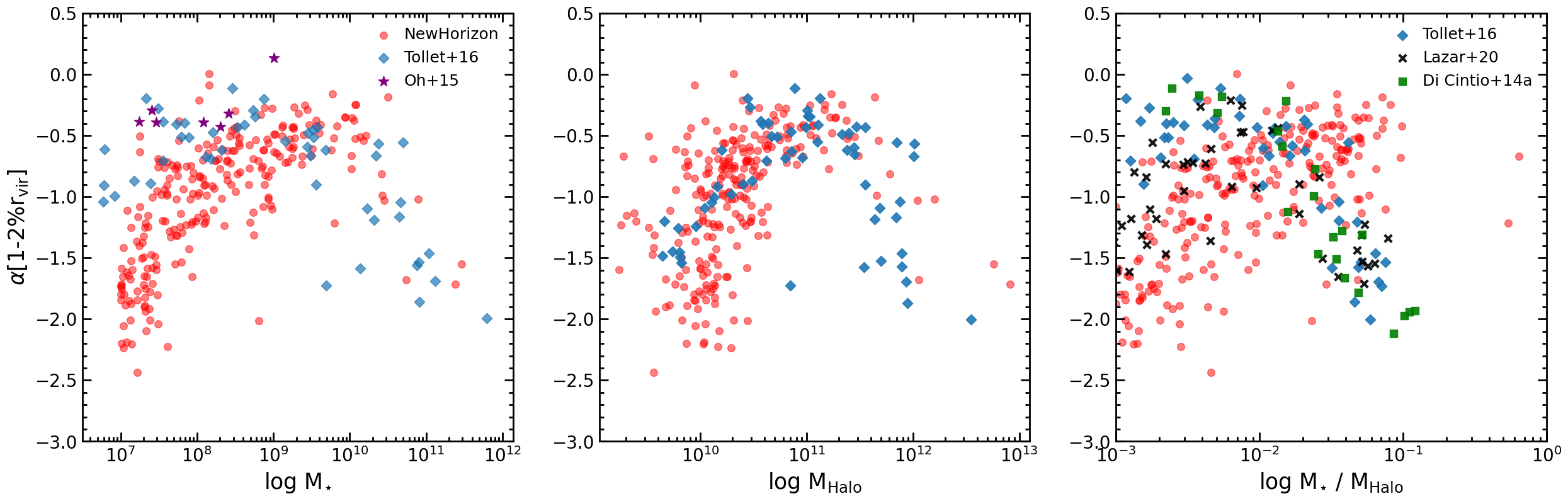}
\caption{DM density profile slope between 1-2$\%$ of the virial radius as a function of the stellar mass (right), halo mass (centre) and stellar/halo mass (right) of galaxies in the full NH (i.e. DM + baryons) run. Galaxies from the NH simulation are shown in red in all the plots. In all panels we compare to the simulation results of \citet{Tollet2016} (shown as blue diamonds) and find reasonable agreement in the stellar and halo mass ranges where cores are found. We compare to observations from \citet{Oh2015} in the left panel (purple stars), finding that, whilst observed cores have similar values for $\alpha$, they are generally found at slightly lower stellar masses than in NH. In the right panel (stellar/halo mass) we also compare to the simulation results of \citet[][green squares]{DiCintio2014} and \citet[][black crosses]{Lazar2020} who find a clear trend between M$_{\star}$/M$_{\rm{Halo}}$ and $\alpha$. We do not see a similar relationship in NH, where M$_{\star}$/M$_{\rm{Halo}}$ increases with $\alpha$.} 
\label{fig:core_comp}
\end{figure*}

It has been shown in previous studies that modern simulations are able to produce DM density profiles that correspond to both cores and cusps when there are baryons present \citep{DiCintio2014,Tollet2016,Chan2015,Dutton2016,Fitts2017,Benitez-Llambay2019,Lazar2020}. In order to investigate this using \texttt{NewHorizon} (NH) we first match halos between a DM-only and full (i.e. DM + baryons) versions of the simulation. This gives us a sample of DM halos with and without baryons for comparison. The matching is confirmed by ensuring that it passes a threshold of sharing a significant number (75\%) of the same DM particles. In addition, if the total mass ratio of the halos in the two runs is outside of the range 0.1 < M$_{\rm{NH}}$/M$_{\rm{NH DM-only}}$ < 10) then it is excluded from our analysis (see \citet{Peirani2017} for more details). With this sample of halos in hand, we begin our analysis by creating DM halo density profiles for all galaxies in both the full NH run and its DM-only counterpart. The volume of NH is large enough that we are able to investigate the DM halos of $\sim$1000 galaxies in the stellar mass range 10$^7$ M$_{\odot}$ < M$_{\star}$ < 10$^{11.5}$ M$_{\odot}$, thus ensuring that we have a statistically significant sample. 

We start by showing, in Figure \ref{fig:sim_comp}, how the M$_{\rm{Halo}}$ vs M$_{\star}$ and M$_{\star}$ vs R$_{\rm{eff}}$ relations in NH compare to those of comparable simulations (APOSTLE, EAGLE CHT10 and NIHAO) and observational results \citep{Behroozi2013,Moster2013}. NH halos of a given mass contain more stellar mass than other simulations (though similar to APOSTLE at lower masses) and also observations although at higher stellar masses are consistent with observations. However the M$_{\star}$ vs R$_{\rm{eff}}$ relation is well matched to the observation data of \citet{Mowla2019}.

In Figure \ref{fig:corecusp_ex} we show two examples of DM halos in the two versions (DM-only and full, i.e. DM + baryons) of NH, which exhibit differing shapes in their central regions. The left panel shows a massive halo (M$_{\rm{halo}}$ = 10$^{12.2}$M$_{\odot}$) which, in the NH full run, contains a galaxy with M$_{\star}$ = 10$^{10.9}$M$_{\odot}$. The DM density profile in both runs shows a clear cusp-like shape with no flattening in the central region. For halos (and galaxies) at such masses the presence of baryons increases the central density of the DM profile, due to the extra baryonic mass creating a deeper gravitational potential. 

The right panel shows a low mass halo (M$_{\rm{halo}}$ = 10$^{10.31}$M$_{\odot}$), which contains a dwarf galaxy with M$_{\star}$ = 10$^{8.15}$M$_{\odot}$ in the NH full run. While the DM-only run shows a cuspy profile, the presence of baryons in this example flattens the central regions in the NH full run, as DM is redistributed within the halo from central regions to larger radii. It therefore seems that, in NH, baryons are able to influence the DM density profiles of halos in multiple ways, depending on the halo/galaxy mass in question. 

\begin{figure}
\centering
\includegraphics[width=0.9\columnwidth]{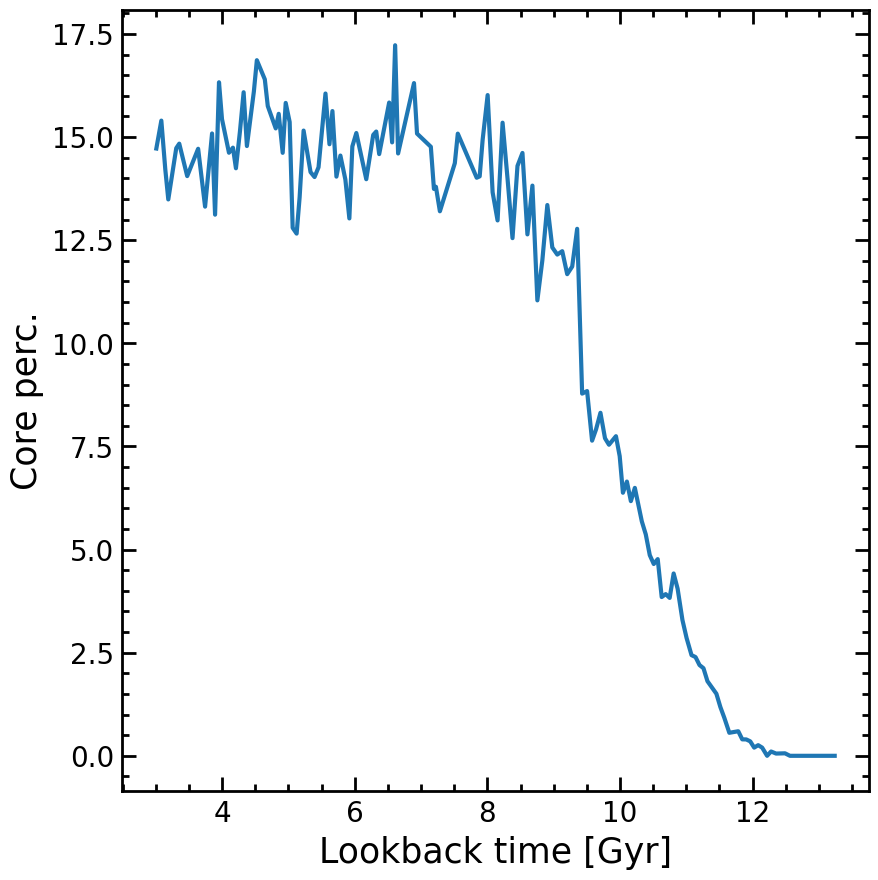}
\caption{The percentage of galaxies that contain a cored DM density profile (defined as $\alpha$ > -0.65) as a function of lookback time in the NH simulation. Initially there are few to no cores in halos but, as star formation starts to increase, this value rises rapidly before remaining roughly constant until the end of the simulation. This suggests that for many galaxies cores are not destroyed and/or cores are constantly being created in galaxies.}
\label{fig:core_perc}
\end{figure}

\begin{figure}
\centering
\includegraphics[width=0.9\columnwidth]{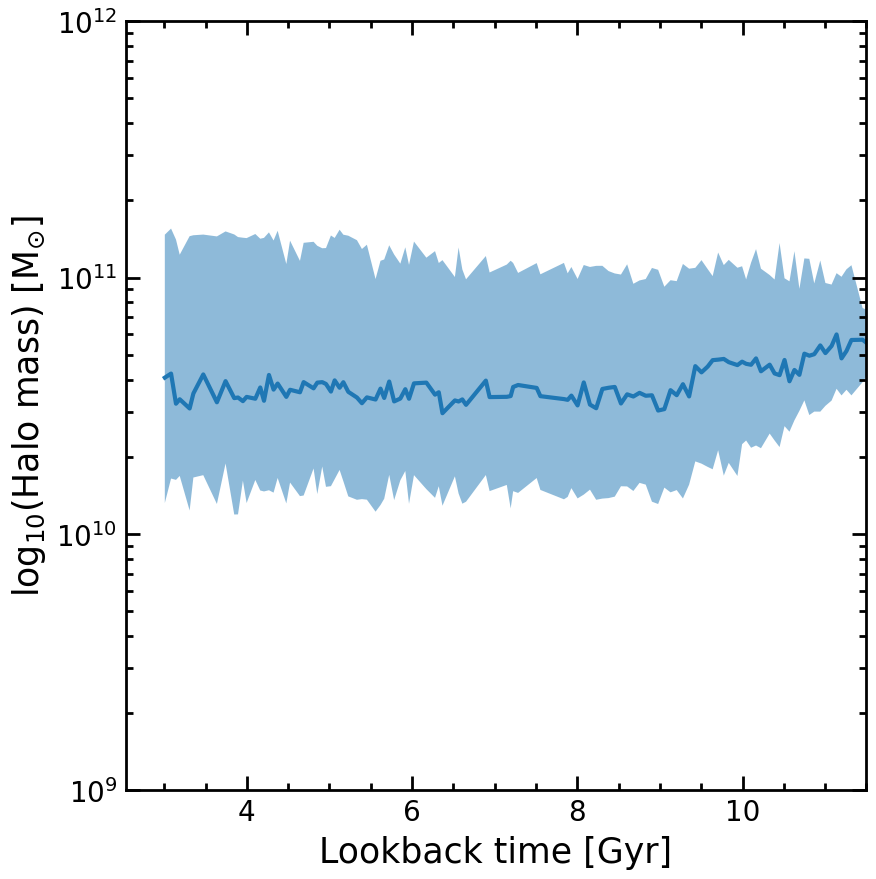}
\caption{The halo mass of galaxies with a cored DM density profile (defined as $\alpha$ > -0.65) as a function of lookback time. The darker line shows the evolution of the median halo mass and the edges of the shaded region represents the 16th and 84th percentiles. The median halo mass remains roughly constant throughout time, indicating that there may be a) a threshold halo mass that allows for the creation and maintenance of a core and b) a secondary threshold above which halos can no longer support a core and so the central regions returns to a cusp.}
\label{fig:core_halomass}
\end{figure}

\subsection{Parametrising DM density profile shapes}
\label{sec:param}

In order to study how baryonic processes affect the inner DM density profiles of our simulated galaxies, we first need to parameterise the profile shapes in these inner regions. To do this we construct the DM density profile by computing the mean density profiles (binned in spherical shells equally spaced in log r) and then fit a power law (C$r^{\alpha}$, where C is a constant) to obtain the slope within a given radial range in log space. In order to compare the results of this study to other work, and to ensure that we probe the same region of each halo regardless of mass, we measure the slope of the profile between 1 and 2$\%$ of a halo's virial radius. We ensure that this is above the DM spatial resolution limit ($\sim$500 pc) and remove any galaxies that do not meet this requirement. It is also important to ensure that our selection of radial range does not influence the overall results. To account for this, we also measure the slope of the DM density profile between 1 and 2 kpc which is usually larger than 2$\%$ of the virial radius. In Appendix \ref{appendix:comp} we show the different slopes obtained for all galaxies using these different radial ranges. As we find that the overall trends do not change with the radial ranges used (even if there is some variation on an individual galaxy basis) we proceed by using the fractional virial radius range for our analysis (i.e. slopes measured within 1 and 2$\%$ of a halo's virial radius).

In Figure \ref{fig:NH_DM-only} we show the inner DM density profile slope ($\alpha$), between $1-2$\% of R$_{\rm{vir}}$ vs the halo mass for all galaxies in the full NH run (colour-coded by their stellar mass) and their corresponding halo in the DM-only run (orange points) at the last timestep of the simulation (which corresponds to z$\sim0.18$). The inner slopes of halos show a clear relationship with halo mass for NH galaxies. In the lowest and highest mass halos we find cusp-like profiles (defined in this study as $\alpha < -0.65$), whereas in intermediate mass halos (with DM masses between 10$^{10.2}$ M$_{\odot}$ and 10$^{11.5}$ M$_{\odot}$) the inner DM profiles are flatter and core-like ($\alpha > -0.65$). We pick this cut-off as no DM-only halos have profiles $\alpha > -0.65$ as well as being roughly consistent with previous studies. At both extremes of halo mass, the halos show evidence of being even denser in their centres compared to their DM-only counterparts, with profiles that are steeper than those predicted by an NFW profile. Halos in the DM-only run are all roughly consistent with the predicted values for an NFW-profile and so are cuspy in nature. Therefore, the presence of baryons is able to cause a halo's centre to be both more or less dense than an NFW-profile, depending on the mass of the halo and the baryonic processes that operate in the central regions. 


\begin{figure*}
\centering
\includegraphics[width=0.9\textwidth]{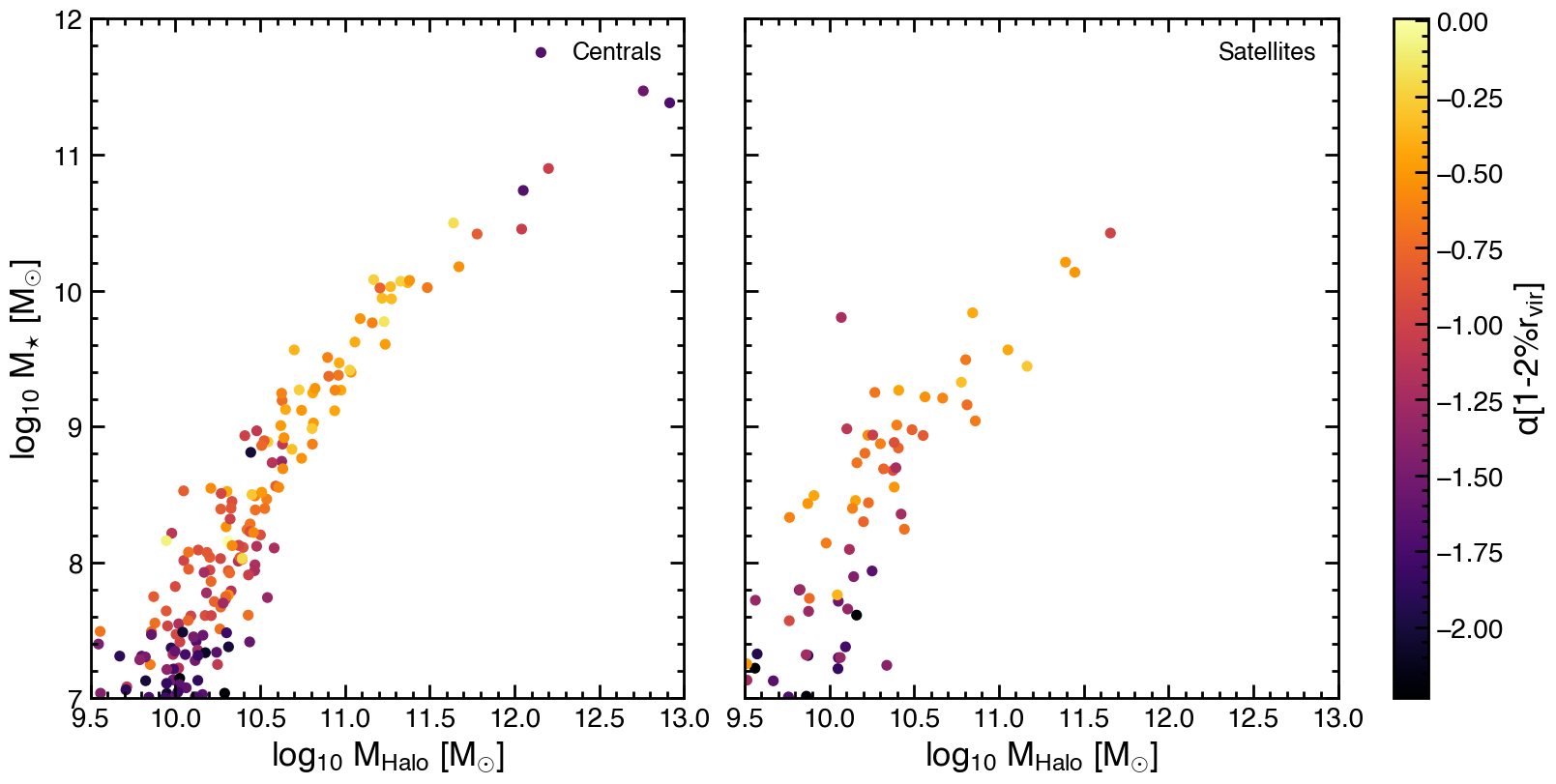}
\caption{Halo mass vs stellar mass for centrals (left) and satellites (right), colour coded by the slope of the DM density profile. Halos with cores are found in both populations which indicates that tidal interactions are not solely responsible for the creation of cores. Cores are found at lower halo and stellar masses in satellites than centrals, likely due to the effect of tidal stripping. }
\label{fig:cent_sat}
\end{figure*}

\begin{center}
\begin{table*}
\centering
\begin{tabular}{| c | c | c | c | c |}
\hline
\hline
Profile & Log stellar mass [M$_{\odot}$] & Log DM mass [M$_{\odot}$] & Log gas mass [M$_{\odot}$] & Log SF gas mass [M$_{\odot}$]\\
& 2 kpc | Total & 2 kpc | Total & 2 kpc | Total & 2 kpc | Total \\
\hline
\hline
Cusp & 7.64 | 8.27 & 8.58 | 10.31 & 8.44 | 9.29 & 6.98 | 6.98\\
Core & 7.80 | 8.56 & 8.57 | 10.37 & 8.47 | 9.41 & 7.83 | 7.97\\
\end{tabular}
\caption{Median values of the masses of different components within galaxies in the mass range 8 < log(M$_{\star}$/M$_{\odot}$) < 9, measured within both the central 2 kpc and the virial radius, for galaxies with cuspy (top row) and cored (bottom row) profiles respectively. The stellar, DM and gas masses are consistent between both profile types, both within 2 kpc and the whole halo. However, the amount of star-forming gas is elevated by an order of magnitude in cored galaxies, indicating that star formation and the abundance of dense, cold gas plays a key role in the formation of cores.}
\label{tab:masses}
\end{table*}
\end{center}

\subsection{$\alpha$ as a function of stellar and halo mass}

Figure \ref{fig:core_comp} shows how the measured $\alpha$ for each halo varies with stellar mass (left panel), halo mass (centre panel) and M$_{\star}$/M$_{\rm{halo}}$ (right panel). The red points indicate the values for galaxies in the full NH run in each panel. In the left panel we compare to the theoretical results of \citet{Tollet2016} and observational results of \citet{Oh2015}. We find reasonable agreement with \citet{Tollet2016} in higher mass galaxies (down to $\sim$10$^9$ M$_{\odot}$), albeit with a larger scatter for NH galaxies. The lowest mass galaxies in NH show significantly cuspier profiles, although the broad transition region where cores are no longer found is consistent. \citet{Oh2015} find cores in a similar stellar mass range, although many cores are at lower stellar mass in their study. It is worth noting, however, that their sample only contains a small number of galaxies. 

When we consider the central panel, we also find agreement between the NH distribution and \citet{Tollet2016} at high and intermediate halo masses. However, below $\sim$10$^{10.2}$ M$_{\odot}$ we find halos with both cores and cusps in NH unlike in \citet{Tollet2016}. In the right panel we consider M$_{\star}$/M$_{\rm{halo}}$, which in many studies is believed to be the key determining factor for core formation. Here we include the simulation results from \citet{Tollet2016}, \citet{DiCintio2014} and \citet{Lazar2020} in our comparison. The simulation studies \citep{DiCintio2014,Tollet2016,Lazar2020} all measure the DM density slope within the same radius, although we note that in \citet{Oh2015} the slope is calculated within a break radius and may therefore be different (see \citet{Oh2011} for more details). Whilst these three studies find a similar relationship between $\alpha$ and M$_{\star}$/M$_{\rm{halo}}$, we find that NH galaxies show a different trend. In NH, cores are found across a range of M$_{\star}$/M$_{\rm{halo}}$ values. Unlike many other studies, which find a dependence on M$_{\star}$/M$_{\rm{halo}}$ and which is deemed to be a proxy of star formation efficiency \citep{DiCintio2014,Tollet2016,Dutton2020}, there is no clear region of the M$_{\star}$/M$_{\rm{halo}}$ relation where galaxies in NH only contain a core. Instead, it appears that once the halo crosses a threshold mass ($\sim$10$^{10.2}$M$_{\odot}$) and enough stars have formed ($\sim$10$^{8}$M$_{\odot}$), there is a possibility a core will be created. The idea of a `threshold halo mass' for core formation to take place has been previously suggested in \citet{Chan2015} but is demonstrated here, using NH, using a much larger galaxy sample. 

As there is no clear trend with M$_{\star}$/M$_{\rm{halo}}$ in NH, unlike previous studies, it would imply that this metric should not be used solely to determine whether a galaxy should contain a core. There is, however, a range of halo and stellar masses where most galaxies have a cored profile. Interestingly, low-mass galaxies in NH may overproduce stars compared to observed abundance matching relations \citep{Dubois2020}, explaining the discrepancy in the $\alpha$-M$_{\star}$/M$_{\rm{halo}}$ relationship. This gives credence to the idea that core formation is not solely dictated by M$_{\star}$/M$_{\rm{halo}}$, as NH creates cores at similar stellar and halo masses to other simulations.

\begin{figure*}
\centering
\includegraphics[width=0.9\textwidth]{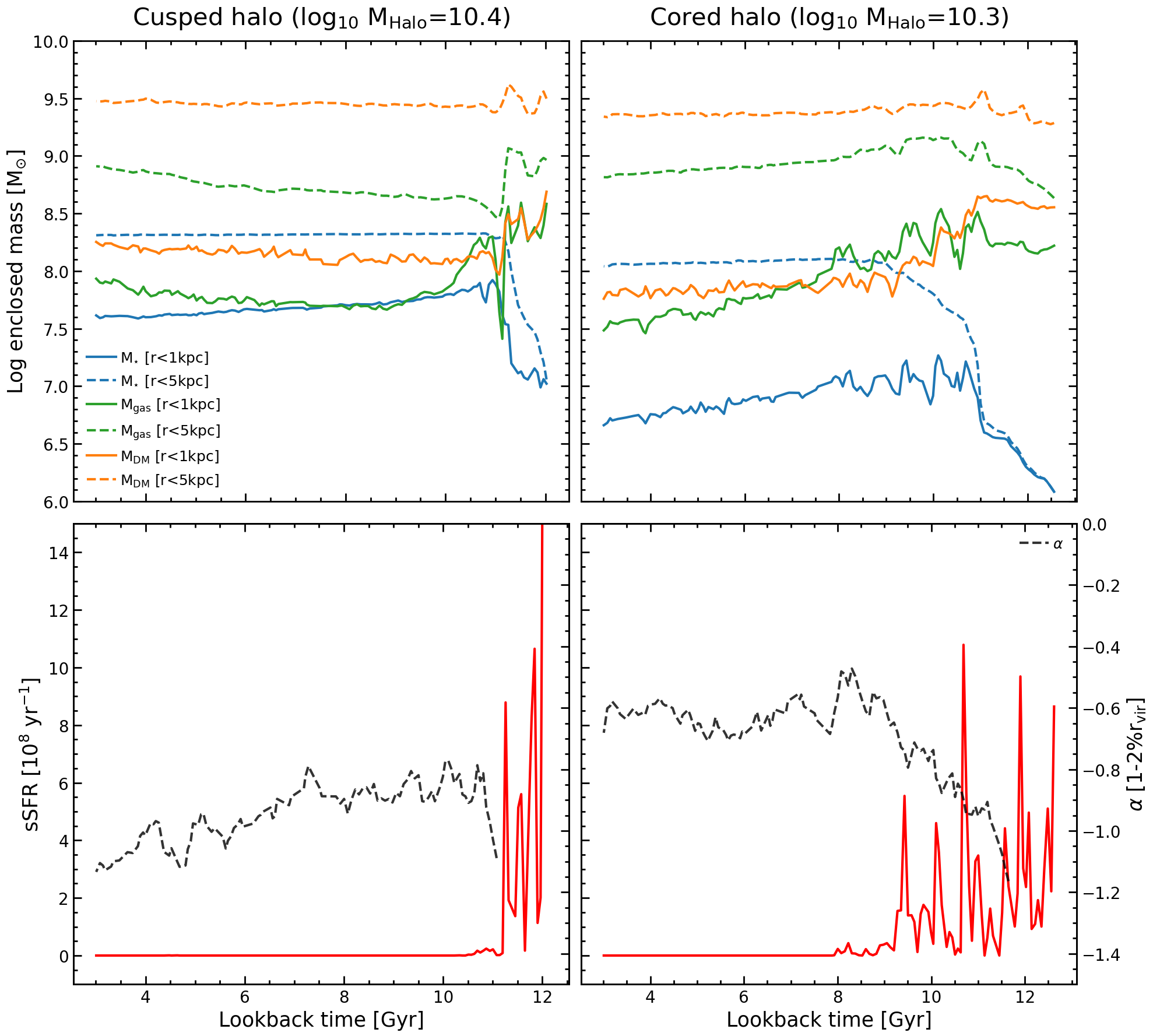}
\caption{\textbf{Top row:} Enclosed mass (gas, DM, stellar) within 1 and 5 physical kpc vs lookback time for a cuspy halo (left) and a cored halo (right). The mass within 1 kpc is shown using the solid lines, while the mass within 5 kpc is shown using the dashed lines. \textbf{Bottom row:} Specific star formation rate of the galaxy (red) and inner DM density profile slope (black dashed line) vs lookback time (note that the slope is only shown when the inner DM density profile is above the resolution criteria set in Section \ref{sec:param}). Cored halos experience bursty star formation over an extended period of time which results in the removal of gas from within the central 1 kpc. When this occurs the DM responds to this removal of mass and also moves out of the central regions. The halo experiences multiple episodes of this process and eventually forms a core. Conversely, low mass cuspy halos have much briefer periods of star formation at high redshift before becoming quenched. This stops the removal of gas from within 1 kpc and means that the gravitational potential is not significantly altered and a core is never formed.}
\label{fig:ratio_core}
\end{figure*}

\begin{figure}
\centering
\includegraphics[width=\columnwidth]{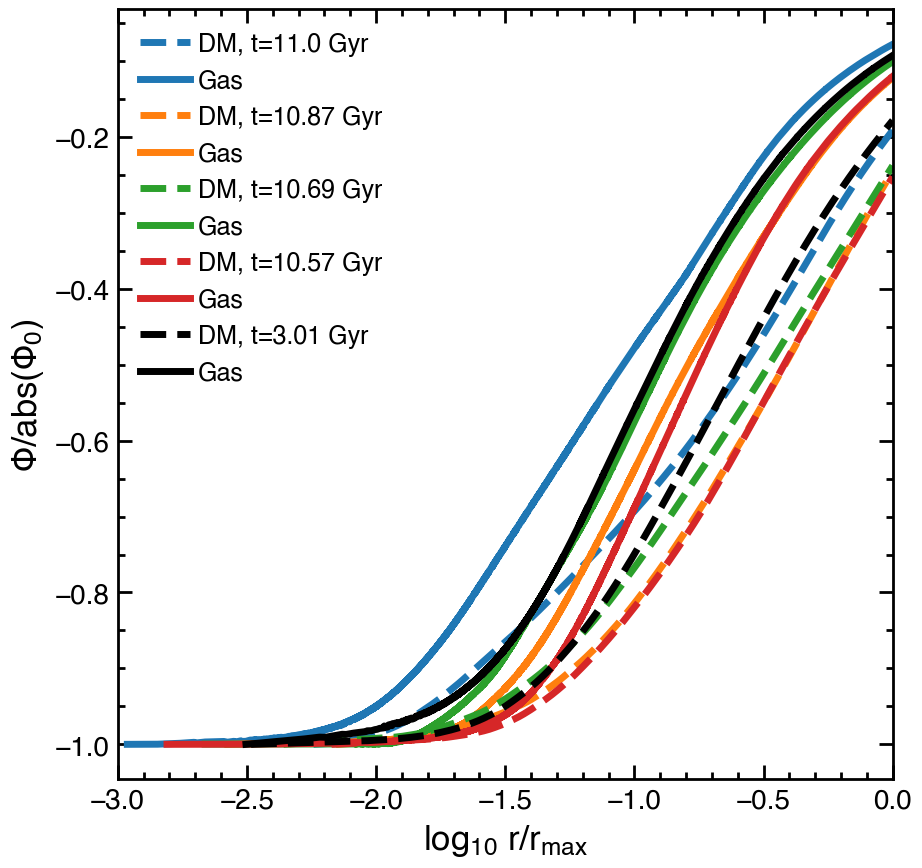}
\caption{The gravitational potential vs log radius, at different look-back times (11, 10.87, 10.69, 10.57 and 3.01 Gyrs), for the halo of a example galaxy that forms a core. Dashed lines indicate the contribution to the potential from the DM whereas solid lines show the gas potential. These look-back times correspond to starbursts that contribute to the creation of the core. All values are normalised by their maximum, in order to compare the changing shape of the potential across cosmic time. We see that even on these short timescales, the bursts of star formation and subsequent movement of gas are able to alter the potential in the central regions of the halo, enabling the formation of a core.}
\label{fig:potential}
\end{figure}

\section{The formation of cores across cosmic time}
\label{sec:CCevo}

A key question in the formation of cores is whether the halo initially forms as a core or whether a core develops due to some physical processes. In this section, we investigate the prevalence of cored DM profiles across cosmic time. Figure \ref{fig:core_perc} shows the percentage of galaxies with core-like profiles (recall that this is defined as $\alpha$ > $-0.65$) as a function of lookback time. We see that the frequency of cores has been relatively constant for the last $\sim$9 Gyrs, suggesting that either, once cores have formed, they are resilient and hard to destroy or cores are constantly being formed to replenish those that are destroyed. At the last timestep we find that $\sim$70$\%$ of cores have existed for at least 3 Gyrs, implying that most cores are relatively long lived. Prior to this period, however, there were few to no cores. Indeed, it is during the period where cosmic star formation starts to increase and then peaks \citep{Madau2014} that the number of cores begins to increase rapidly. This not only allows time for halos to grow to a mass capable of maintaining a core but also suggests that star formation plays a key role in core formation. 

As we now see that cores have existed across most of cosmic time, we can further assess the concept of a `threshold halo mass', by looking at the mass of halos which contain cores throughout the simulation. In Figure \ref{fig:core_halomass} we show the median halo mass of galaxies with cores throughout the simulation (with the edges of the shaded region indicating the 16th and 84th percentile values). The median halo mass of cores is relatively constant across time at $\sim$10$^{10.3}$ M$_{\odot}$. This implies that the process that creates the flattened density profile requires some mass limit to be reached and also suggests that there is an upper limit beyond which the halo centres will return to NFW-like profiles again, as cores are not found at all halo masses. This, combined with the results from Figure \ref{fig:core_comp}, suggests that cores will form once halos are both above a given mass and when star formation is occurring within the galaxy. Given that this halo mass is roughly consistent with the other simulations we have compared our results to \citep{DiCintio2014,Tollet2016,Lazar2020}, it appears that disparate baryonic prescriptions can give rise to core formation within different simulations. 


\subsection{How and why do cores form?}

In order to study how and why cored profiles form, we consider a region of the stellar and halo mass parameter space where halos have both cored and cuspy inner profiles and examine the physical differences between galaxies in this regime. We proceed by selecting galaxies with stellar masses in the range 8 < log(M$_{\star}$/M$_{\odot}$) < 9 in the final snapshot ($z=0.18$) as this contains a large number of both cores and cusps, whilst also retaining enough resolution in the simulation to study the evolution of the galaxies. 


We start by looking at external processes that could create cored profiles. Tidal interactions between galaxies have been shown to strongly affect the DM distribution of galaxies, albeit usually at the outskirts \citep{Jackson2021a}. Figure \ref{fig:cent_sat} shows halo mass versus stellar mass for central galaxies (left) and satellites (right) defined by the AdaptaHOP structure finder, colour-coded by their values of $\alpha$. Both central and satellite galaxies contain a variety of core and cusp profiles, with cores in satellites being found at slightly lower halo masses (likely due to tidal stripping of the outskirts reducing their original halo mass). This suggests that becoming a satellite does not initially alter the central DM distributions of galaxies and therefore that tidal processes are not a likely cause of core creation, at least not directly. This also indicates that it is possible to find galaxies with cores as satellites of larger galaxies, such as the those in the local group, as has been previously suggested by \citet{Walker2011}.

As external processes are unlikely to be able to alter the central part of DM profiles (at least until the outskirts have been stripped), we proceed by investigating the internal physical properties of galaxies in this transition region. In Table \ref{tab:masses}, we show the median values of stellar, DM, total gas and star-forming gas mass, measured both within the central 2 kpc and within the virial radius, for all cored and cuspy galaxies in our transition region. The stellar, DM and total gas mass are consistent, regardless of the shape of the DM density profile, with gas and DM being the dominant contributors to the central mass budget in both cases. However, in the cored profiles the amount of star-forming gas ($n_{\rm{H}}>$10 H cm$^{-3}$) is significantly elevated in the centre, suggesting that there is (or has recently been) increased star formation in these galaxies compared to their cuspy counterparts. Feedback processes from star formation are known to be able to drive outflows of gas which, if contributing a significant amount of mass, can alter a galaxy's gravitational potential and induce the DM to move to larger radii \citep{Navarro1996,Read2005,Mashchenko2006,Pontzen2014,Martizzi2013}.


Therefore, the relative contributions to the total mass in the central regions of the halo is key to determining the shape of the profile. In order to understand how the mass budget impacts the DM profile, we compare the central mass evolution of two example core and cusp galaxies. Both galaxies are centrals and, by the end of the simulation, are hosted in a similarly sized halos (log M$_{\rm{Halo}}$$\sim$10.3). Figure \ref{fig:ratio_core} shows the enclosed mass within two radial distances (1 kpc and 5 kpc) as a function of lookback time for a galaxy with a cuspy halo (left panel) and a cored halo (right panel). These galaxies represent typical examples of cuspy and cored halos in this transition regime. The blue curves show the stellar mass evolution, the green curves represent total gas and the orange curves show DM, with solid lines indicating the mass within 1 kpc and dashed lines showing the mass within 5 kpc. In the bottom panels, the red lines show the evolution of the specific star formation rate (sSFR) and the dashed black line shows the evolution of the inner DM density slope of the galaxies in question.

At large lookback times (i.e. in the early Universe) both galaxies show highly elevated sSFRs, which coincides with a rapid increase in stellar mass in the central regions. At the same epoch, gas and DM constitute roughly equal amounts of mass within 1 kpc and are dominant over the stellar component. In the case of the galaxy that ends up with a cuspy halo, the gas component drops rapidly and star formation quenches, leaving the DM component at a constant level and the dominant constituent of the central mass budget.

However, in the case of the galaxy that ends up with a cored halo, we see repeated bursts of star formation, with the gas mass within 1 kpc repeatedly rising and falling, with the falls occurring shortly after the peak sSFR values. When looking at all cores in NH, we find that no galaxy has a core when bursty star formation lasts for less than 2 Gyrs. The gas and DM components within 1 kpc are similar in mass initially but, unlike in the cuspy case above, the gas mass remains high and even becomes the dominant central mass component at $8 < t < 10$ Gyr. Despite star formation taking place, the amount of stellar mass being formed is less than the oscillations in gas mass, and thus the removal of gas from the centre is a combination of stars forming and gas being moved to larger radii. 

As the gas mass is comparable to the DM mass, moving large amounts of gas to larger radii will alter the gravitational potential, allowing the DM to migrate to larger radii as well. For example, in the right panel of Figure \ref{fig:ratio_core}, as gas is removed from the centre (at around 11 Gyrs), the DM mass within 1 kpc also decreases. Although gas in-falls to the centre again and the mass increases the DM mass does not and remains relatively constant. Gas is then once again removed leading to a further decrease in the DM. In this way the DM mass is removed in stages by bursts of star formation and gas removal, which flattens the central DM density profile and creates a core. We can also calculate the dynamical time for the halo during this period, within 1 kpc, to understand how this compares to the oscillations of gas. In the example of the core in Figure \ref{fig:ratio_core}, we find that when the core is forming, the dynamical time is 0.2 Gyrs, which is roughly equal to the duration of the observed bursts.

In order for a core to be formed in this manner, the gravitational potential of the galaxy needs to be altered from a classical NFW profile. As previously shown, large amounts of gas could be moved from the very central regions to larger radii, thus altering the shape of the potential. In Figure \ref{fig:potential} we show the gravitational potential of the example cored galaxy as a function of its radius. In order to compare the shape of the potential at different times, as well as to its DM-only counterpart, we normalise the potential by its lowest value ($\Phi$$_{0}$) and the halo's radius by its maximum value (r$_{\rm{max}}$). We show both the DM component of the potential (dashed lines) and the gas component (solid lines) for different look-back times that correspond to moments before and after bursts of star formation which create the core and the final timestep of the simulation (black). Between the first two timesteps shown, the potential of both the DM and gas is significantly altered due to the removal of gas (and subsequently DM) from the centre of the halo. After this period the potential starts to return to its previous shape until another burst of SF further alters it towards a core. Indeed, even by the end of the simulation the potential has not returned to its original shape, before the bursts of star formation occurred.

\begin{figure}
\centering
\includegraphics[width=\columnwidth]{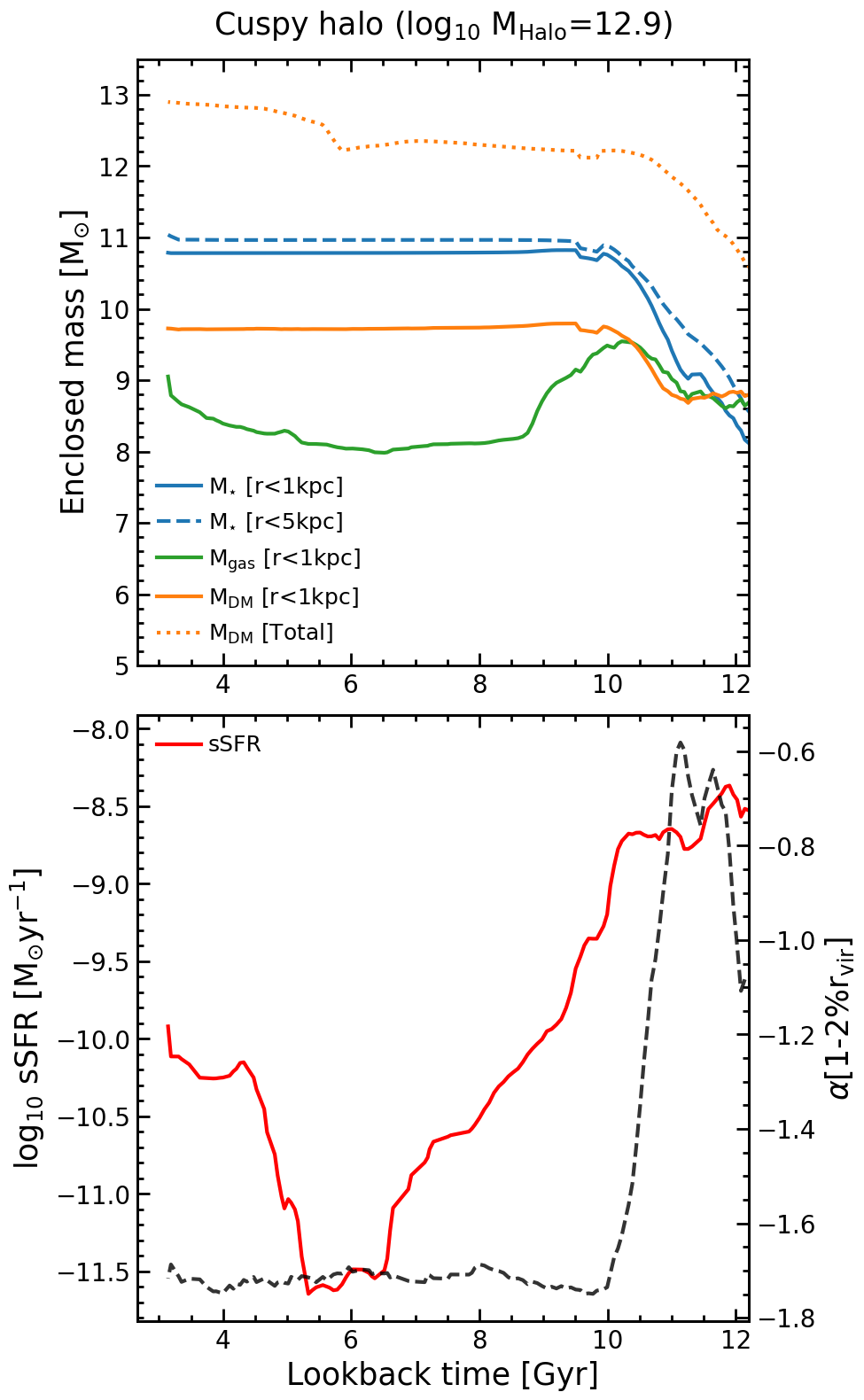}
\caption{\textbf{Top panel:} The enclosed mass (within 1 kpc) for M$_{\star}$ (blue), M$_{\rm{gas}}$ (green) and M$_{\rm{DM}}$ (orange) vs look-back time (Gyr) for a massive halo (M$_{\rm{Halo}}$=10$^{12.9}$ M$_{\odot}$) with a cusp at the final timestep of the full NH (i.e. DM + baryons) run. We also show the enclosed mass within 5 kpc for M$_{\star}$ (blue dashed line) to explore whether stars are being formed in-situ or not. \textbf{Bottom panel:} The evolution of the sSFR (red) with lookback time and $\alpha$ (black dashed line) for this galaxy. At large look-back times, this halo had a cored profile when the gas and DM masses were roughly equal in the centre. However, ongoing star formation with a high sSFR changes the makeup of the central masses. As stars begin to dominate the mass in the centre of the halo, the core disappears as a cusp is reformed.}
\label{fig:ratio_cusp}
\end{figure}

\subsection{The evolution of massive galaxy profiles - the reformation of cusps}

As seen in Figure \ref{fig:core_comp}, above a certain halo mass most galaxies in the simulation have cuspy DM density profiles. However, if core formation is determined purely by a threshold halo mass and star formation, have these galaxies had cores in their past and, if so, what removed them? In Figure \ref{fig:ratio_cusp}, we show an example of a massive galaxy, containing a cusp at the final timestep in NH, with a stellar and halo mass of 10$^{11.3}$M$_{\odot}$ and 10$^{12.9}$M$_{\odot}$ respectively. In the top panel we show the stellar (blue), gas (orange) and DM (green) mass enclosed within the inner 1 kpc (and the stellar mass within the inner 5 kpc as a blue dashed line) as a function of look-back time. For the lower panel we show the sSFR (red line) of the galaxy as a proxy for SN feedback and the DM density slope within $1-2$\% of R$_{\rm{vir}}$ (black dashed line).


This massive galaxy experiences a period when it has a cored DM density profile at high redshift, coinciding with high sSFR and comparable gas/DM/stellar masses in the central 1 kpc. However, unlike in Figure \ref{fig:ratio_core}, the DM density profile steepens again as the stellar mass continues to increase in both the inner 1 and 5 kpc, despite ongoing star formation and high gas mass. The stellar mass increases in both radial bins, which suggests that it is increased central in-situ star formation (rather than existing mass falling into the centre or ex-situ mass accretion) that drives the reformation of the cusp. Unlike in the cored example (Figure \ref{fig:ratio_core}), sustained star formation leads to the stellar component becoming dominant over both gas and DM in the central regions. This increased stellar component deepens and anchors the potential, reducing the outflow of gas and thus stopping the dynamical heating of DM particles. The DM reacts to this new, deeper potential removing the core and recreating the cusp. Therefore the halo's inability to control star formation is at least partly responsible for the rejuvenation of the cusp. 

In addition, just like in the cored example, we calculate the dynamical time for the halo within 1 kpc. We see a similar dynamical time ($\sim$0.2 Gyrs) whilst this halo has a core. However, as the star formation continues the dynamical time rapidly decreases to 0.04 Gyrs meaning that the cusp can rapidly reform, as is observed in the slope of the DM density profile.

The potential role of BHs (and therefore AGN) is also worth considering, especially as it has been previously shown to affect DM density profiles in larger galaxies \citep{Peirani2017}. However, for galaxies in the mass ranges where cores are found in NH the BH occupation fraction is $\sim$30$\%$. In addition, in those galaxies which contain BHs both cuspy and cored halos are found. Therefore it is logical to conclude that, at least in this mass range, BHs and AGN are not primarily responsible for the creation of cores or the reformation of cusps. 




\section{Summary and conclusions}
\label{sec:summary}

Dwarf galaxies dominate the galaxy number density in all environments across cosmic time. However, due to their low stellar masses, and consequently fainter surface brightnesses, they have remained difficult to study in past surveys. This, in  part, has contributed to many tensions between observations and theoretical predictions within $\Lambda$CDM in the dwarf regime. One longstanding tension is the core-cusp problem. DM-only simulations suggest that DM halos should have a cuspy profile, with the density continuing to increase towards the centre. However, many observational studies of dwarf galaxies indicate that they actually possess flattened density profiles in their central regions, known as cores. 

Here, we have used the \texttt{NewHorizon} (NH) cosmological simulation, in conjunction with its DM-only counterpart, to examine the evolution of the DM density profiles of galaxies across a large range of stellar masses and environments (from the field to large groups). This has enabled us to study the diversity of DM density profiles in both dwarf and massive galaxies in a statistical fashion, for the first time within a cosmological volume. Our main conclusions are as follows:

\begin{itemize}

    \item The central DM density slope ($\alpha$) between $1-2$\% of R$_{\rm{vir}}$ can be used to probe whether a halo contains a core or a cusp. In our study we consider galaxies with $\alpha$ > -0.65 as having cores. While halos in NH exhibit DM density profiles with both cores and cusps, only cusps are present in halos in the DM-only run. Thus, it is the presence of baryons that drives the observed diversity of DM density profiles. 
    
    \item Cores in NH form when there are repeated bursts of star formation, in a galaxy that has a central gas mass that is comparable to the DM mass. No galaxy (at least in NH) creates a core when this bursty star formation lasts for a period shorter than 2 Gyrs. The resultant SN feedback removes gas from the central regions, altering the shape of the (central) gravitational potential. The DM responds to this change and moves to larger radii. This process is typically repeated multiple times, with the central DM being removed in stages, until star formation ceases. If this process occurs for a sufficient length of time ($\sim$2-3 Gyrs), the shape of the DM density profile is eventually transformed from an initial cusp into a core.

    \item Conversely cusps can be reformed if the central star formation continues rapidly after the core forms. In this scenario a core forms, albeit briefly, before being removed when stellar mass becomes dominant in the most central regions. This mass comes from central in-situ star formation rather than stellar mass infall from larger radii into the central regions or ex-situ accretion (such as mergers). At this stage, stars dominate the central gravitational potential and, since they are not directly affected by SN feedback, anchor the new potential in place. This, in turn, causes the DM to adiabatically cool, reforming a cuspy DM density profile. 


    \item Halos that contain galaxies in the upper (M$_{\star}$ $\ge$ 10$^{10.2}$ M$_{\odot}$) and lower (M$_{\star}$ $\le$ 10$^{8}$ M$_{\odot}$) ends of the stellar mass distribution contain cusps. However, in halos that contain galaxies that have intermediate (10$^{8}$ M$_{\odot}$ $\le$ M$_{\star}$ $\le$ 10$^{10.2}$ M$_{\odot}$) stellar masses, $\alpha$ becomes more positive than in the DM-only run (i.e. the profiles become core-like), with this becoming most common in halos with DM masses between 10$^{10.2}$ M$_{\odot}$ and 10$^{11.5}$ M$_{\odot}$. 
    
    \item The evolution of $\alpha$ with stellar and halo mass in NH is broadly consistent with what is seen in other comparable simulations (except at the lowest stellar masses) and also in limited available observational data (given the difficulties in inferring a DM density profile from observations). However, the trend of increasing core frequency with increasing M$_{\star}$/M$_{\rm{Halo}}$, that is observed in some other simulations, is not seen in \texttt{NewHorizon}, indicating that this may not be the best metric to determine whether a galaxy has a core or a cusp (as has been previously suggested). 

    \item The fraction of galaxies that contain a core becomes roughly constant ($\sim$15 per cent) after a redshift of $z\sim1.5$, with cores not forming until the cosmic star formation rate increases around a redshift of $z\sim2$. The mass of halos that contain a core is roughly constant in the mass range 10$^{10.2}$ M$_{\odot}$ < M$_{\star}$ < 10$^{11}$ M$_{\odot}$, with a median value of 10$^{10.3}$ M$_{\odot}$, which indicates (1) a potential threshold halo mass below which cores cannot form and (2) a maximum halo mass above which cores are removed. Additionally at the final timestep 70$\%$ of cores have existed for at least 3 Gyrs, suggesting that most cores are relatively long-lived.
     
\end{itemize}

We conclude that, in NH, a cosmological simulation that adopts a $\Lambda$CDM cosmology, the core-cusp problem is not a challenge to the standard paradigm. While this has been shown in previous theoretical work, we are able to perform this test with a statistically significant number of galaxies, over a larger range of galaxy masses, and within a cosmological volume. The limited amount of observational data makes comparisons to the real Universe difficult, though this may be addressed in future deep-wide surveys, which will allow us to constrain the stellar masses where cores form in real galaxies. As NH has been shown to successfully recreate cores and cusps, it is a unique tool to investigate other `small-scale' challenges to $\Lambda$CDM, such as the diversity of rotation curves in dwarf galaxies, which will be explored in a forthcoming paper (Jackson et al. in prep). 


\section*{Acknowledgements}
SK acknowledges support from the STFC [grant numbers ST/S00615X/1 and ST/X001318/1]. SK also acknowledges a Senior Research Fellowship from Worcester College Oxford. S.K.Y. acknowledges support from the Korean National Research Foundation (NRF-2020R1A2C3003769, NRF-2022R1A6A1A03053472). S.K.Y acted as a corresponding author as the head of the group whilst most of this investigation was being performed at Yonsei University. This research has made use of the Horizon cluster on which the simulation was post-processed, hosted by the Institut d'Astrophysique de Paris. We warmly thank S.~Rouberol for running it smoothly. This work is partially supported by the grant
Segal ANR-19-CE31-0017 of the French Agence Nationale de la Recherche and by the National Science Foundation under Grant No. NSF PHY-1748958.
This work was granted access to the HPC resources of CINES under the allocations c2016047637, A0020407637 and A0070402192 by Genci and as a “Grand Challenge” project granted by GENCI on the AMD Rome extension of the Joliot Curie supercomputer at TGCC. Part of NewHorizon was performed using Nurion at KISTI under the allocation of KSC-2017-G2-0003, and the large data transfer was supported by KREONET which is managed and operated by KISTI. S.K.Y. acknowledges support from the Korean National Research Foundation (NRF-2020R1A2C3003769, NRF-2022R1A6A1A03053472). This study was funded in part by the NRF-2022R1A6A1A03053472 grant. TK was supported by the National Research Foundation of Korea (Nos. 2020R1C1C1007079 and 2022R1A6A1A03053472)

For the purpose of open access, the authors have applied a Creative Commons Attribution (CC BY) licence to any Author Accepted Manuscript version arising from this submission. 

\section*{Data Availability}
 All data used in this paper can be made available upon reasonable request to the first author.



\bibliographystyle{mnras}
\bibliography{bib}



\appendix
\section{Radius range comparison}
\label{appendix:comp}
\begin{figure}
\centering
\includegraphics[width=\columnwidth]{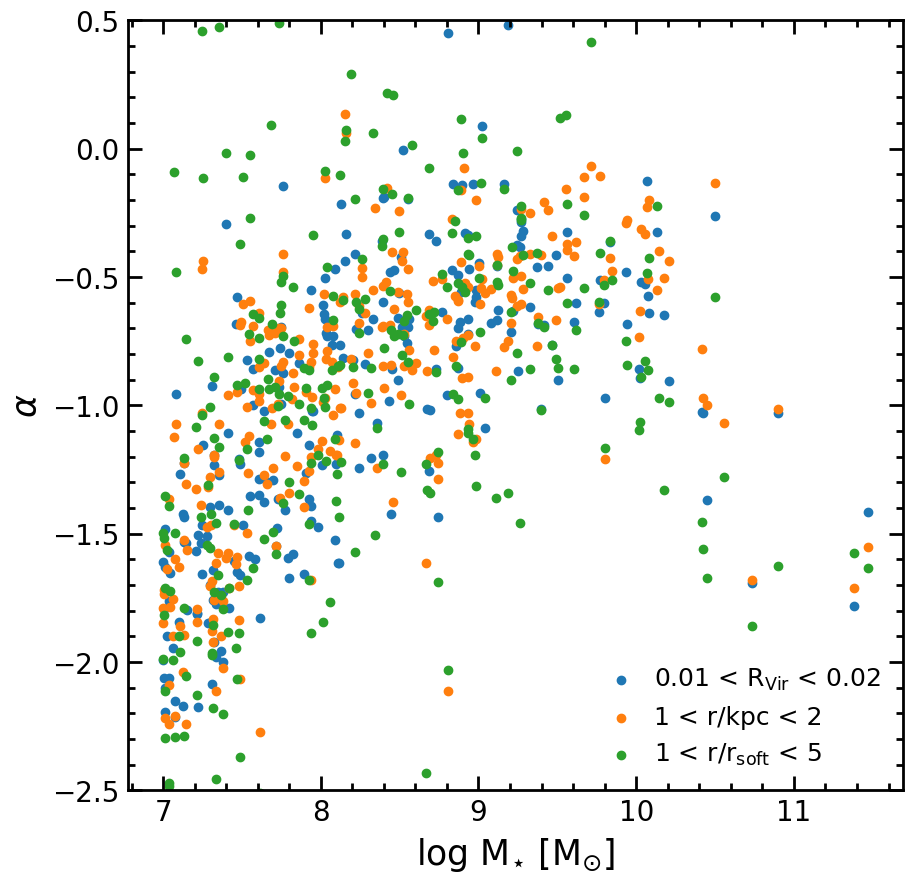}
\caption{Inner DM density profile slope ($\alpha$) vs stellar mass. We calculate the slope of the inner DM density profile using three different radial ranges: 0.01 < r/R$_{\rm{Vir}}$ < 0.02 (blue), 1 < r/kpc < 2 (orange) and 1 < r/r$_{\rm{soft}}$ < 5 (orange) where r$_{\rm{soft}}$ is $\sim$500 pc. All three radial ranges produce consistent values for $\alpha$, suggesting that the results are relatively independent of which range is selected.}
\label{fig:alpha_comp}
\end{figure}

In Figure \ref{fig:alpha_comp} we compare the slope of the inner DM density profiles using three different radial ranges as a function of stellar mass. In blue we show profiles calculated in the range 0.01 < R$_{\rm{vir}}$ < 0.02. The orange points show the same for 1 < r/kpc < 2, while the green points show profiles calculated in the range 1 < r/r$_{\rm{soft}}$ < 5, where r$_{\rm{soft}}$ is $\sim$500 pc. For most galaxies, the radius at which the inner DM density profile gradient is measured does not alter the results found. The overall trend remains the same, with cores formed in galaxies between 10$^8$ M$_{\odot}$ < M$_{\star}$ < 10$^{10.2}$ M$_{\odot}$ and cusps formed in the very low and very high mass galaxies. This also suggests that the cores produced are larger than these radii, indicating that many cores in NH are at least 2 kpc in size. 


\bsp	
\label{lastpage}
\end{document}